\documentclass[aps,pra,twocolumn,showpacs]{revtex4} 
\newcommand{\ket}[1]{|#1\rangle} 
\newcommand{\bra}[1]{\langle #1|} 
 
\usepackage{graphicx} 
\begin{document} 
\title{Effect of intersubsystem coupling on the geometric phase 
in a bipartite system} 
\author{X.X. Yi$^1$\footnote{Electronic address: yixx@dlut.edu.cn}, 
and Erik  Sj\"oqvist$^2$\footnote{Electronic address: eriks@kvac.uu.se}} 
\affiliation{$^1$Department of physics, Dalian University of 
Technology, Dalian 116024, China\\ 
$^2$Department of quantum Chemistry, Uppsala University, Box 518, 
Se-751 20 Uppsala, Sweden} 
\begin{abstract} 
The influence of intersubsystem coupling on the cyclic adiabatic 
geometric phase in bipartite systems is investigated. We examine 
the geometric phase effects for two uniaxially coupled 
spin$-\frac{1}{2}$ particles, both driven by a slowly rotating 
magnetic field. It is demonstrated that the relation between the 
geometric phase and the solid angle enclosed by the magnetic field 
is broken by the spin-spin coupling, in particular leading to a 
quenching effect on the geometric phase in the strong coupling 
limit.  
\end{abstract} 
\maketitle 
The geometric phase, originally conceived by Berry \cite{berry84} for 
cyclic adiabatic evolution of pure quantal states, has been extensively 
studied \cite{shapere89,thouless83} and generalized, for example 
to nonadiabatic evolution \cite{aharonov87}, mixed states 
\cite{uhlmann86,sjoqvist00a}, and open systems \cite{gaitan98}. 
The appearance of the geometric phase in composite entangled systems 
with no intersubsystem coupling has been analyzed for a pair of 
entangled spins in a time-independent uniform magnetic field 
\cite{sjoqvist00b} and for the case of entangled spin 
pairs in a rotating magnetic field \cite{tong03a}; it has also 
attracted interest in connection to polarization-entangled photon 
pairs \cite{hessmo00}, the topology of the SO(3) rotation group 
\cite{milman03}, Bell's theorem \cite{bertlmann03}, as well as in 
relation to the mixed state geometric phases of the subsystems 
\cite{tong03b}. 
 
The importance of geometric phases to robust control of 
quantal systems, such as in fault tolerant quantum computation 
\cite{zanardi99}, has triggered extension of the geometric 
phase for composite systems to interacting subsystems. This 
topic has been addressed for two isolated interacting spins 
\cite{sjoqvist00b}, for systems entangled with a quantized 
driving field \cite{fuentes02}, in relation to unitary 
representations of quantum channels \cite{ericsson03}, 
and more recently for a pair of interacting spin$-\frac{1}{2}$ 
particles with one of the spins driven by a slowly rotating 
magnetic field \cite{yi04}. Concern about the effect of 
interaction on the geometric phase may also arise in the 
application of the geometric phase to systems with intra-variable 
couplings, such as, e.g., spin-orbit coupling in 
atomic systems, where the entanglement among a distinguished 
set of observables becomes an attractive issue in quantum 
information processing \cite{barnum03}. 
 
In this work, we develop the theory geometric phases of bipartite
systems with intersubsystem coupling undergoing adiabatic cyclic
evolution. We examine the effect of intersubsystem coupling on the
pure state geometric phase of the composite system, as well as on the
geometric phases associated with the reduced density operators of the
subsystems. We calculate and analyze the geometric phases in the case
of two uniaxially interacting spin-$\frac{1}{2}$ (qubit) systems with
the same magnetic dipole moment and driven by a slowly precessing
magnetic field; a case of relevance to, e.g., entanglement creation by
adiabatic passage techniques \cite{unanyan01} as well as to NMR 
quantum computation \cite{gershenfeld97,laflamme02}. Finally, we 
briefly discuss a possible extension of this analysis to systems with
intra-variable coupling.
 
Let a quantal system $S$ be exposed to the Hamiltonian $H(Q)$, $Q$ 
being some external control parameters. Suppose that the Hilbert space 
${\cal H}$ of $S$ is $N$ dimensional and that $Q$ varies around a 
closed path ${\cal C}: t\in [0,T] \rightarrow Q_t$ in parameter space, 
so that $H(Q_T)=H(Q_0)$. Expansion of the solution of the Schr\"{o}dinger 
equation in the instantaneous eigenstates $\ket{n(Q_t)}$ of $H(Q_t)$ 
yields 
\begin{eqnarray} 
\ket{\Psi(t)} = \sum_{n=1}^N c_n (t) \ket{n(Q_t)} . 
\end{eqnarray} 
If $T$ is large enough, the adiabatic theorem \cite{messiah62} entails 
that transitions between the instantaneous energy eigenstates are 
negligible, making $|c_n|$ approximately time-independent, so that 
the final state reads ($\hbar = 1$ from now on) 
\begin{equation} 
\ket{\Psi (T)} = \sum_{n=1}^N |c_n| e^{-i \int_0^T E_n (t) dt} 
e^{i\gamma_n [{\cal C}]} \ket{n(Q_0)} , 
\end{equation} 
where $E_n (t)$ and $\gamma_n [{\cal C}]$ are the instantaneous 
nondegenerate energy eigenvalue and the cyclic adiabatic geometric 
phase, respectively, associated with the $n$th energy eigenstate. 
 
Now, let us focus on the case where $S$ has a natural bipartite 
decomposition in terms of subsystems $S_a$ and $S_b$ with corresponding 
Hilbert spaces ${\cal H}_a$ and ${\cal H}_b$. For such 
$S$, the energy eigenvectors may be put on Schmidt form 
\begin{equation} 
\ket{n(Q,g)} = \sum_{k=1}^N \sqrt{p_k^{(n)}(Q,g)} 
\ket{a_k^{(n)}(Q,g)} \otimes \ket{b_k^{(n)}(Q,g)} , 
\end{equation} 
where the Schmidt vectors $\ket{a_k^{(n)}(Q,g)} \otimes 
\ket{b_k^{(n)}(Q,g)}$ are characterized by $\langle a_k^{(n)}(Q,g) 
\ket{a_l^{(n)}(Q,g)} = \delta_{kl}$ and $\langle b_k^{(n)}(Q,g) 
\ket{b_l^{(n)}(Q,g)} = \delta_{kl}$, $g$ is some set of fixed coupling 
parameters, and $N=\min \big( \dim {\cal H}_a , 
\dim {\cal H}_b \big)$. The Schmidt decomposition is unique provided 
the nonvanishing coefficients $p_k^{(n)}$ are nondegenerate, i.e., 
$p_k^{(n)} \neq p_l^{(n)}, \ \forall k,l$. The geometric phase 
associated with the path ${\cal C}$ in parameter space may be 
written as 
\begin{eqnarray} 
\gamma_{ab}^{(n)} [{\cal C};g] & \equiv & \gamma_n [{\cal C}] = 
i \oint_{{\cal C}} dQ \cdot \bra{n(Q,g)} \nabla_Q \ket{n(Q,g)} 
\nonumber \\ 
 & = & 
\sum_k \Big( \widetilde{\Gamma}_{a;k}^{(n)} [{\cal C};g] + 
\widetilde{\Gamma}_{b;k}^{(n)} [{\cal C};g] \Big) , 
\label{eq:general} 
\end{eqnarray} 
where we have used that the $p_k^{(n)}$'s sum up to unity. Here, 
\begin{eqnarray} 
\widetilde{\Gamma}_{\xi;k}^{(n)} [{\cal C};g] = 
i \oint_{{\cal C}} dQ \cdot \bra{\widetilde{\xi}_k^{(n)}(Q,g)} 
\nabla_Q \widetilde{\xi}_k^{(n)}(Q,g) \rangle 
\end{eqnarray} 
are geometric phases of the weighted Schmidt vectors 
$\ket{\widetilde{\xi}_k^{(n)}(Q,g)} = \big[ p_k^{(n)}(Q,g) \big]^{1/2} 
\ket{\xi_k^{(n)}(Q,g)}$ pertaining to subsystem $S_{\xi=a,b}$. 
 
Next, let us introduce the concept of {\it nontransition eigenstates}. 
These are defined as energy eigenstates where the $p_k$'s are 
time-independent. For vanishing intersubsystem coupling, i.e., when 
$g=0$, only such states occur since the time evolution operator then 
takes the bi-local form $U_{ab} = U_a \otimes U_b$, which exactly 
preserves the Schmidt coefficients $p_k^{(n)}$.  On the other hand, 
transitions usually occur in the presence of coupling and the 
nontransition condition is in general only valid for specific 
paths. Closed paths of this kind are rare but have been found 
and studied for a spin-spin interaction model in Ref. \cite{yi04}. 
 
Suppose that there exists a nontransition state tracing out a 
closed path ${\cal D}$ in parameter space. For such a path, we 
may compute the cyclic geometric phase as 
\begin{eqnarray} 
\gamma_{ab}^{(n)} [{\cal D};g] & = & i\sum_k p_k^{(n)} (Q_0,g) \Big( 
\Gamma_{a;k}^{(n)} [{\cal D};g] + \Gamma_{b;k}^{(n)} [{\cal D};g] \Big) , 
\nonumber \\ 
\label{eq:nontransition} 
\end{eqnarray} 
where 
\begin{eqnarray} 
\Gamma_{\xi;k}^{(n)} [{\cal D};g] & = & 
i\oint_{{\cal D}} dQ \cdot \bra{\xi_k^{(n)}(Q,g)} \nabla_Q 
\xi_k^{(n)}(Q,g) \rangle 
\nonumber \\ 
 & = & \big[p_k^{(n)} (Q_0,g)\big]^{-1} 
\widetilde{\Gamma}_{\xi;k}^{(n)} [{\cal D};g] 
\label{schmidtv} 
\end{eqnarray} 
with $\xi=a,b$, constitute the one-particle geometric phases of the 
Schmidt vectors. The nontransition property makes it natural to extend 
Ref. \cite{sjoqvist00a} and define mixed state geometric phases for 
the two interacting subsystems as 
\begin{equation} 
\gamma_{\xi}^{(n)} [{\cal D};g] = 
\arg \sum_k p_k^{(n)}(Q_0,g) 
\exp \Big( i \Gamma_{\xi;k}^{(n)} [{\cal D};g] \Big) . 
\label{eq:subgp} 
\end{equation} 
Thus, the geometric phases of the subsystems are taken as the average 
of phase factors pertaining to the nontransition eigenstates of the 
reduced density operators, weighted by the corresponding eigenvalues. 
 
Now, consider two qubits $a$ and $b$ as represented by a pair of 
spin-$\frac{1}{2}$ particles coupled by a uniaxial exchange 
interaction in the $z$ direction. In the presence of a time-dependent 
external magnetic field ${\bf B}(t) =B_0 \hat{{\bf n}}(t)$ with 
the unit vector $\hat{{\bf n}}=(\sin\theta \cos\phi, 
\sin\theta \sin\phi,\cos\theta)$, the Hamiltonian of this 
system reads \cite{abragam61} 
\begin{equation} 
H(t) = 4J S_a^z S_b^z + 
\mu {\bf B}(t) \cdot ({\bf S}_a+{\bf S}_b) 
\label{ha1} 
\end{equation} 
with $\mu$ the magnetic dipole moment assumed to be equal for the 
two spins. The first part of the Hamiltonian describes the exchange 
interaction (spin-spin coupling) with coupling constant $J$, $\mu$ 
is the gyromagnetic ratio, and ${\bf S}_{\xi} = (S_{\xi}^x,S_{\xi}^y, 
S_{\xi}^z)$ is the $\xi$th spin operator ($\xi=a,b$). An explicit 
physical scenario for this model could be NMR 
experiments on $^{13}C$-labeled trichloroethylene in which the nuclear 
spins of the two $^{13}C$ nuclei could act as two spin$-\frac{1}{2}$ 
systems with nearly the same magnetic dipole moment (the chemical 
shift of the two nuclei for this system is typically a fraction 
$10^{-5}$ of their precession frequency \cite{laflamme02}). In terms of 
the orthonormalized total spin eigenstates $\ket{S;M}$, $S=0,1$ and 
$M=-S,\ldots,S$, in the $z$ direction and in units of $\mu B_0$, 
the Hamiltonian Eq.~(\ref{ha1}) can be expressed in the block-matrix 
form \cite{unanyan01} 
\begin{equation} 
H(t)=\left( \matrix{ H_c(t) & 0  \cr 
 0 & -g \cr } \right),\label{ha2} 
\end{equation} 
where 
\begin{equation} 
H_c(t)=\left( \matrix{ g-\cos\theta & \frac{1}{\sqrt{2}}\sin\theta 
e^{i\phi} &0 \cr \frac{1}{\sqrt{2}}\sin\theta e^{-i\phi} & -g 
&\frac{1}{\sqrt{2}}\sin\theta e^{i\phi}\cr 0 & \frac 
{1}{\sqrt{2}}\sin\theta e^{-i\phi} &g +\cos\theta \cr } \right)  
\label{ha3} 
\end{equation} 
with rescaled coupling constant $g=\frac {J}{\mu B_0}$ that may be
controlled by changing the magnitude of the external magnetic field.
Thus, the spin singlet $\ket{0;0}$ is decoupled from the triplet
states $\ket{1;M}$, $M=-1,0,1$. Therefore, in contrast to the case
where only one of the subsystems interacts with ${\bf B} (t)$
\cite{yi04}, the adiabatic geometric phase acquired by the singlet
state vanishes here.
 
First, let us consider the case of vanishing spin-spin coupling 
characterized by $g=0$. In this case, the total spin in the 
$\hat{\bf n}$ direction commutes with $H(t)$ so that the adiabatic 
geometric phases may be expressed in terms of the total spin 
projection quantum number $M$ along the magnetic field and the 
solid angle $\Omega$ enclosed by the path ${\cal D}$ in parameter 
space. For $M = \pm 1$ we obtain 
\begin{eqnarray} 
\gamma_{ab}^{(\pm 1)} [{\cal D};0] & = & \mp \Omega , 
\nonumber \\ 
\gamma_{a}^{(\pm 1)} [{\cal D};0] & = & 
\gamma_{b}^{(\pm 1)} [{\cal D};0] = \mp \frac{\Omega}{2} , 
\label{g0} 
\end{eqnarray} 
where the former follows from the standard Berry formula $-M\Omega$ 
\cite{berry84}. The $M = 0$ eigenstates are two-fold degenerate 
and $\ket{\Psi^{(0)}} = \alpha \ket{1;0} + \beta \ket{0;0}$ for 
any complex numbers $\alpha$ and $\beta$ is an energy eigenstate. 
While $\gamma_{ab}^{(0)} [{\cal D};0]$ vanishes when taking 
$\ket{\Psi^{(0)}}$ around ${\cal D}$, the corresponding mixed 
state geometric phase for the two subsystems become 
\begin{eqnarray} 
\gamma_{a}^{(0)} [{\cal D};0] & = & 
- \gamma_{b}^{(0)} [{\cal D};0] 
\nonumber \\ 
 & = & 
- \arctan \left[ 2\textrm{Re} \big( \alpha^{\ast} \beta \big) 
\tan \left( \frac{\Omega}{2} \right)\right] 
\end{eqnarray} 
provided $2\textrm{Re} \big( \alpha^{\ast} \beta \big) \neq 0$. On the 
other hand, when $2\textrm{Re} \big( \alpha^{\ast} \beta \big) = 0$, which 
for example occurs if the two spins are associated with 
indistinguishable entities, in case of which the singlet and 
triplet states cannot mix, the reduced density operators of the 
subsystems are degenerate, and the corresponding geometric phases 
become undefined since no direction in space is singled out by the 
corresponding Bloch vectors. In all other cases, we have 
$\gamma_{ab}^{(M)} [{\cal D};0] = \gamma_{a}^{(M)} 
[{\cal D};0] + \gamma_{b}^{(M)} [{\cal D};0]$, which is due to the 
spherical symmetry of the model in the $g=0$ case. 
 
For $g\neq 0$, the spherical symmetry is broken, and there is no 
simple relation neither between the geometric phases and the solid 
angle nor between the geometric phase of the composite system 
and those of the subsystems. Thus, to proceed we need to diagonalize 
$H_c (t)$ to obtain its eigenstates as ($n=-,0,+$) 
\begin{eqnarray} 
\ket{\Psi^{(n)}} & = & e^{i\phi} A^{(n)}(\theta,g) 
\ket{1;-1} + B^{(n)}(\theta,g) 
\ket{1;0} 
\nonumber \\ 
 & & + e^{-i\phi} C^{(n)}(\theta,g) \ket{1;1} 
\label{eigenv} 
\end{eqnarray} 
with 
\begin{eqnarray} 
A^{(n)} & = & 
\frac{1}{\sqrt{2M^{(n)}}} 
\big[ X^{(n)} - \cos\theta \big] \sin\theta , 
\nonumber\\ 
B^{(n)} & = & \frac{1}{\sqrt{M^{(n)}}} \big[ (X^{(n)})^2 - 
\cos^2\theta \big] , 
\nonumber\\ 
C^{(n)} & = &  \frac{1}{\sqrt{2M^{(n)}}} 
\big[ X^{(n)} + \cos\theta \big] \sin\theta , 
\nonumber \\ 
M^{(n)} & = & 
(X^{(n)})^4 + (1 - 3\cos^2\theta) (X^{(n)})^2 + \cos^2\theta . \ \ 
\label{eigenf} 
\end{eqnarray} 
The shifted instantaneous energy eigenvalues $X^{(n)} = 
X^{(n)}(\theta,g) \equiv E_n-g$ of the Hamiltonian 
Eq. (\ref{ha3}) are solutions of 
\begin{eqnarray} 
X^3+2gX^2-X-2g\cos^2\theta = 0, 
\end{eqnarray} 
which yields $X^{(\pm)}(\theta,g)=\pm \cos\theta$ and 
$X^{(0)}(\theta,g)=-2g$ in the limit of $g\rightarrow \infty$. The 
Schmidt coefficients read 
\begin{eqnarray} 
p_1^{(n)} & = & 1-p_2^{(n)} = 
\frac{1}{2} \Big( 1+ \left( A^{(n)} + C^{(n)} \right) 
\nonumber \\ 
 & & \times \sqrt{2(B^{(n)})^2 + \big(C^{(n)} - A^{(n)} 
\big)^2} \Big) 
\nonumber \\ 
 & = & \frac{1}{2} \Big( 1+r^{(n)} (\theta,g) \Big) . 
\end{eqnarray} 
Thus, $p_1^{(n)}$ and $p_2^{(n)}$ are determined by the $\phi$ 
independent effective Bloch vector $r^{(n)}$ and it follows that the 
nontransition paths are those where $\theta$ is constant.  For closed 
paths ${\cal D} : t\in [0,T] \rightarrow (\phi_t,\theta_t) = 
(2\pi t/T,\theta)$ with $r^{(n)} \neq 0$ and $T$ large, we uniquely 
obtain the adiabatic geometric phases for the corresponding Schmidt 
vectors pertaining to subsystem $S_{\xi=a,b}$ as 
\begin{eqnarray} 
\Gamma_{\xi,1}^{(n)} [{\cal D},g] = 
-\Gamma_{\xi,2}^{(n)} [{\cal D},g] = 
- \pi \left( 1- F^{(n)} \cos\theta \right) , 
\label{eq:geoschmidt} 
\end{eqnarray} 
where the scale factor 
\begin{eqnarray} 
F^{(n)} & = & F^{(n)} (\theta,g) 
\nonumber \\ 
 & = & \frac{\sin\theta}{\sqrt{(X^{(n)})^4 - 
2(X^{(n)})^2 \cos^2 \theta + \cos^2 \theta}} 
\end{eqnarray} 
comprises the effect of intersubsystem coupling on the 
geometric phase of the Schmidt vectors. The geometric phases 
of the composite system and those of the subsystems read 
\begin{eqnarray} 
\gamma_{ab}^{(n)} [{\cal D},g] & = & 
-2\pi r^{(n)} \Big( 1- F^{(n)} 
\cos\theta \Big) , 
\nonumber \\ 
\gamma_{\xi}^{(n)} [{\cal D},g] & = & 
-\arctan \Big[ r^{(n)} \tan \Big( \pi 
\big[ 1 - F^{(n)} \cos\theta \big] \Big) \Big] 
\nonumber \\ 
\end{eqnarray} 
with $\xi = a,b$. 
 
\begin{figure} 
\includegraphics*[width=0.9\columnwidth, 
height=0.95\columnwidth]{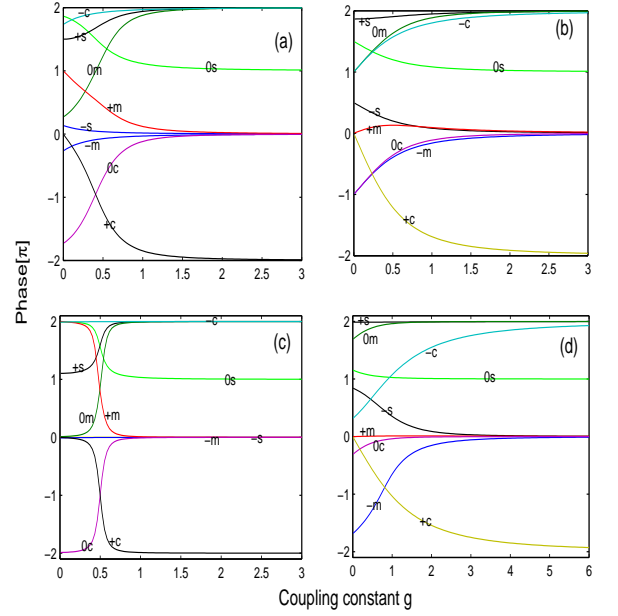} \caption{An illustration of the
geometric phases of the composite system (nc), the sum of the
corresponding mixed state geometric phases for the two qubits (nm),
and the geometric phase pertaining to one of the Schmidt vectors (ns),
corresponding to the three instantaneous eigenvectors
$\ket{\Psi^{(n)}}$, where $n=-,0,+$ and $\ket{\Psi^{(n)}}$ become the
spin triplet states $\ket{1;-1},\ket{1;0}, \ket{1;1}$ in the strong
spin-spin coupling limit. The horizontal axes show the rescaled
dimensionless coupling constant $g$. The composite system is assumed
to undergo an adiabatic evolution in one of the instantaneous
eigenstates, such that the corresponding Schmidt coefficients are
time-independent. These states are characterized by constant polar
angles (a) $\theta = \pi/6$, (b) $\theta=\pi/3$, (c) $\theta=\pi/30$
(close to zero) and (d) $\theta=9\pi/20$ (close to $\pi/2$).}
\label{fig1}
\end{figure} 
 
The dependence of the geometric phase upon the coupling constant $g$ 
is illustrated in Fig. \ref{fig1}. An interesting feature of 
Fig. \ref{fig1} is that all geometric phases for the composite system 
and its subsystems tend to an integer multiple of $2\pi$ when $g 
\rightarrow \infty$. This limit corresponds to the case when the 
second term in the Hamiltonian in Eq. (\ref{ha1}) can be ignored. In 
other words, the triplet states $\ket{1;M}$ would become the 
instantaneous eigenstates of the system with $g\rightarrow \infty$, 
thus making the enclosed area in state space to vanish and thereby 
the geometric phase factors become trivial. It is worth stressing 
that the Hamiltonian in Eq. (\ref{ha1}) has permutation symmetry, 
resulting in $\gamma_a^{(n)}[{\cal D};g] = \gamma_b^{(n)}[{\cal D};g]$.  
In fact, the mixed state geometric phases presented in Fig. \ref{fig1} 
are for $\gamma_a^{(n)}[{\cal D};g] + \gamma_b^{(n)}[{\cal D};g] = 
2\gamma_a^{(n)}[{\cal D};g] = 2\gamma_b^{(n)}[{\cal D};g]$; 
the sum of mixed state geometric phases of the two qubits being 
equal to twice the geometric phase for each subsystem. Note in 
particular that $\gamma_{ab}^{(n)} [{\cal D};g] \neq \gamma_a^{(n)} 
[{\cal D};g] + \gamma_b^{(n)}[{\cal D};g]$ in general. The 
state of the composite system are entangled at most time when 
$g\neq 0$, this indicates that there are at least two nonvanishing 
Schmidt coefficient $p_k$ and thus most the geometric phases for the 
Schmidt vector in Eq. (\ref{schmidtv}) would have similar properties 
in the limit $g\rightarrow \infty$ except that pertaining to 
$|\Psi^{(0)}\rangle$, where $X^{(0)}(\theta, g\rightarrow \infty) 
\rightarrow -g$, and then $F^{(0)}$ tends to zero, consequently, 
$\Gamma_{\xi,2}^{(0)}[{\cal D},g\rightarrow \infty]\rightarrow \pi$ 
as shown by 0s lines in Fig. \ref{fig1}. 
 
\begin{figure} 
\includegraphics*[width=0.95\columnwidth,height=0.6\columnwidth]{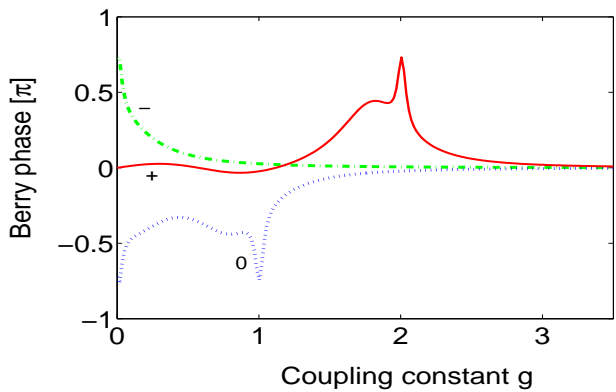} 
\caption{The geometric phase in units of $\pi$ for the eigenstates 
$|\Psi^{(\pm)}\rangle$ and $|\Psi^{(0)}\rangle$ {\it vs.} the rescaled 
dimensionless coupling constant $g$ with $\theta=\pi\sin(\pi t/T)$, 
$T \gg 1$ being the period of precession chosen in accordance with 
the requirement of adiabaticity.} \label{fig2} 
\end{figure} 
 
For transition paths ${\cal C}$ along which the polar angle $\theta$ 
and thereby the Schmidt coefficients vary, the geometric phase of 
the composite system reads 
\begin{eqnarray} 
\gamma_{ab}^{(n)} [{\cal C},g] & = & 
-\oint_{{\cal C}} r^{(n)} \Big( 1- F^{(n)} 
\cos\theta \Big) d\phi . 
\end{eqnarray} 
In Fig. 2, $\gamma_{ab}^{(n)} [{\cal C},g]$ is shown as a function 
of the rescaled coupling constant $g$ for the closed path ${\cal 
C}: t\in[0,T] \rightarrow (\phi_t,\theta_t) = (\pi t/T,\pi \sin 
(\pi t/T))$. With $g \rightarrow \infty$, the phases tend to zero, 
in analogy with the nontransition case discussed above. 
 
An interesting extension of above analysis is to the case of systems
with intra-variable coupling. For example, one may consider an atom
with electronic orbital and spin angular momentum ${\bf L}$ and ${\bf
S}$, respectively, precessing in a time-dependent magnetic field ${\bf
B}$, the Hamiltonian describing such a system reads, $H = \mu {\bf
n}(t) \cdot \big( {\bf L} + 2{\bf S} \big) + g {\bf L} \cdot {\bf S}$,
the last term describes the spin orbit coupling. An analysis is 
expected to show that the spin-orbit coupling would affect the 
geometric phase of the atom in a similar way as in the intersubsystem 
coupling case.
 
In conclusion, we have analyzed the cyclic adiabatic geometric phase 
of bipartite systems, focusing on the effect of intersubsystem 
coupling. We have distinguished two different kind of evolution 
in regard to whether or not the Schmidt coefficients are time-dependent. 
The geometric phases of the subsystems naturally extends in terms of 
the standard mixed state geometric phase \cite{sjoqvist00a} in the 
nontransition case, i.e., when the Schmidt coefficients and thereby 
the eigenvalues of the corresponding reduced density operators are 
fixed. We have found a striking evidence for a strong influence of the 
intersubsystem coupling on the geometric phases for two uniaxially 
coupled spin$-\frac{1}{2}$ systems, such as a divergence from the 
standard relation to the solid angle enclosed by the driving magnetic 
field, leading to a quenching effect on the geometric phases in the 
strong coupling limit. This latter result has also been demonstrated 
in the transition case, where the Schmidt coefficients are changing 
around the curve in parameter space. Physically, the quenching effect 
may be viewed as a consequence of the broken spherical symmetry as 
expressed by the existence of a preferred direction in space singled 
out by the uniaxial exchange direction of the spins, making the 
energy eigenstates independent of the slowly rotating magnetic field.  
\vskip 0.3 cm 
X.X.Y. acknowledges financial support from EYTP of M.O.E, and NSF of
China (Project No. 10305002). E.S. acknowledges financial support from
the Swedish Research Council.
 
\end{document}